\documentclass[aps,pra,superscriptaddress,reprint,floatfix,notitlepage,balancelastpage]{revtex4-1}
\usepackage{graphicx,amsmath,amssymb,bm,color,engord}
\usepackage[usenames,dvipsnames]{xcolor}
\usepackage[colorlinks,colorlinks,urlcolor=Magenta,citecolor=Magenta,linkcolor=black]{hyperref}

\begin{document}

\title{Trapped imbalanced fermionic superfluids  in one dimension: A variational approach}

\author{Kelly R. Patton}
\email[\hspace{-1.4mm}]{kpatton1@ggc.edu}
\affiliation{School of Science and Technology, Georgia Gwinnett College, Lawrenceville, Georgia 30043, USA}
\author{Dominique M. Gautreau}
\altaffiliation[Present address: ]{School of Physics and Astronomy, University of Minnesota, Minneapolis, MN 55455, USA}
\author{Stephen Kudla}
\author{Daniel E. Sheehy}
\email[\hspace{-1.4mm}]{sheehy@lsu.edu}
\affiliation{Department of Physics and Astronomy, Louisiana State University, Baton Rouge, Louisiana 70803, USA}

\date{February 9, 2017}
\begin{abstract} We propose and analyze a variational wave function
for a population-imbalanced one-dimensional Fermi gas that
allows for 
Fulde-Ferrell-Larkin-Ovchinnikov (FFLO) type pairing correlations among the two fermion species, while
also accounting for the harmonic confining potential. 
%
%
%
In the strongly interacting regime, we find large
spatial oscillations of the order parameter, indicative of an
FFLO state. 
%
The
obtained density profiles versus imbalance 
are consistent with recent experimental  results as well as with theoretical
calculations based on combining Bethe ansatz with the local
density approximation.  Although we find no signature of the FFLO
state in the densities of the two fermion species, we show that the oscillations of
the order parameter appear in density-density correlations,  both
in-situ and after free expansion. Furthermore, above a critical polarization, the value of which depends on the interaction, we find the unpaired Fermi-gas state to be energetically more favorable. 
\end{abstract}
\maketitle

\section{Introduction} After the tremendous success of the
Bardeen-Cooper-Schrieffer  (BCS)  theory of superconductivity \cite{BCS}
interest quickly turned towards the possibility of  other more exotic
forms of superconductivity (or fermionic superfluidity).   One of the first 
theoretical proposals for a novel superconductor was the 
 Fulde-Ferrell-Larkin-Ovchinnikov (FFLO) \cite{FF,LO}  state, predicted
to occur in systems with   mismatched Fermi energies, i.e.,~a spin-$\uparrow$ and
spin-$\downarrow$ population imbalance.   The FFLO state is predicted to 
occur, for a conventional three-dimensional $s$-wave superconductor, very close to the 
so-called Chandrasekhar-Clogston (CC) limit \cite{Clogston,Chandrasekhar}, which is
the point where the energy penalty due to the Fermi energy mismatch is larger
than the energy gained from pairing.

 The resulting first order phase transition from the BCS phase to an imbalanced
normal phase that occurs at the CC limit has been well established experimentally
in both electronic superconductors in an externally applied magnetic field and also
cold fermionic atomic gases under an imposed population imbalance.  However, 
the FFLO state, which is theoretically predicted to occupy a region of the phase diagram
close to the CC limit, has never been observed.  

 Unlike the BCS state,  the
Cooper pairs of the FFLO state exhibit a spatial variation in the local pair amplitude. In a
homogeneous system the corresponding FFLO wavevector ${\bm Q}$ is approximately equal to the
difference of the  Fermi surface wave vectors $|{\bm Q}|\simeq k_{{\rm F}\uparrow}-k_{{\rm F}\downarrow}$
for the majority (denoted by spin-$\uparrow$) and minority (spin-$\downarrow$) fermion species.  
 The pairing amplitude takes the form $\Delta({\bf r}) \propto {\rm e}^{i{\bm Q}\cdot {\bf r}}$ in the  FF phase
and $\Delta({\bf r}) \propto \cos{(\bm Q}\cdot {\bf r})$ in the LO phase, with the latter believed to be more stable than
the former.  
%
%
Unfortunately, experimental verification has been lacking, and although some 
evidence of the FFLO state has been seen in cold atom~\cite{LiaoNature2010} and condensed matter~\cite{MayaffreNatPhys,Prestigiacomo} settings, 
no signature of the periodic modulation of the order parameter has
been observed to date.

The regime of stability of the FFLO state, as a function of parameters such as interaction strength
and population imbalance, is theoretically predicted to be
strongly dependent on  dimensionality, with the regime of stability smallest in
three dimensions ~\cite{RS2010,Parish2007}, becoming larger for two spatial
dimensions~\cite{Conduit,HeZhuang,RV,Leo2011,Wolak2012,LevinsenBaur,Caldas,ParishLevinsen,Yin,SheehyFFLO,Toniolo} 
and largest in one dimension~\cite{Orso2007,Hu2007}.
Although many experiments on imbalanced Fermi gases have investigated the three-dimensional 
regime~\cite{Zwierlein2006,Partridge2006,Shin2006,Partridge2006prl,Navon2010,Olsen2015} (finding 
no evidence of the FFLO state), recent experiments have explored Fermi gases in  one~\cite{LiaoNature2010,Revelle2016} or two spatial 
dimensions~\cite{Sommer2012,Zhang2012,Ries,Murthy,Boettcher,Fenech,Cheng2016,Mitra2016}
 using an appropriate trapping potential.

\begin{figure}
\vspace{0.4cm}
\includegraphics[scale=0.66]{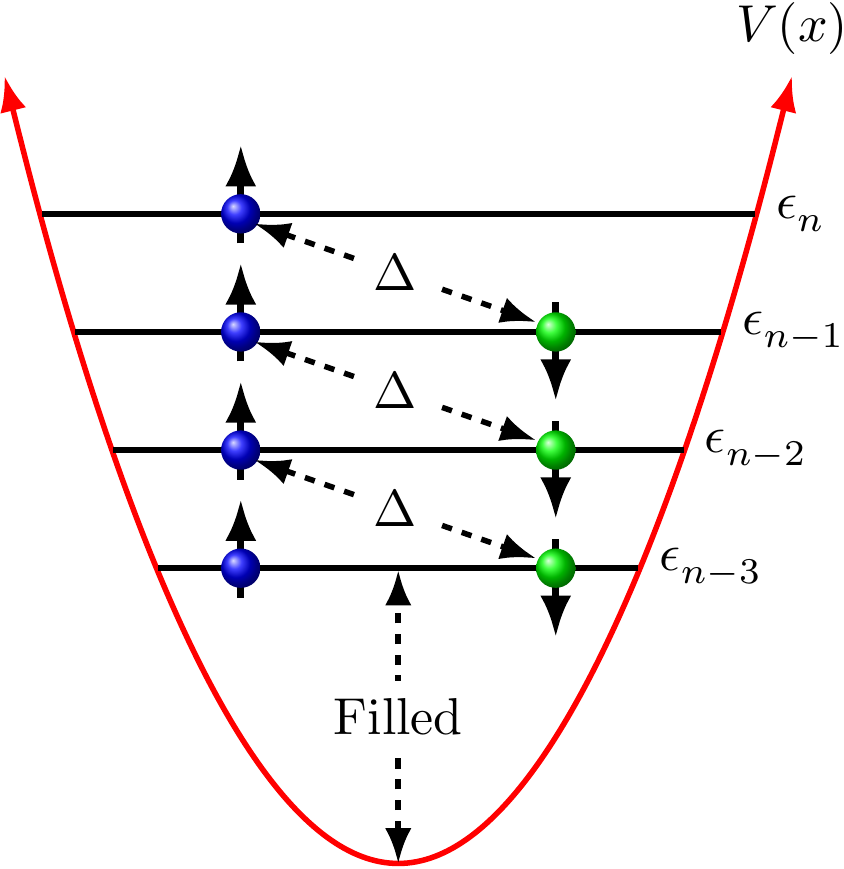}%
\caption{\label{fig1} Schematic of the FFLO wave function for a
trapped one-dimensional gas Eq.~\eqref{FFLO wave function} with
a population imbalance given by $q = N_\uparrow - N_\downarrow$.  Spin-$\uparrow$
fermions at harmonic oscillator level $n$ are paired with spin-$\downarrow$ fermions at level
$n-q$ (our illustration is for the case of $q=1$.)  The excess spins-$\uparrow$ occupy low-energy
levels with $0\leq n\leq q-1$.  
 }
\end{figure} 

Here, our major interest in the one-dimensional (1D) regime, which has been the subject of
extensive recent theoretical work.  
The  theoretical work investigating  the FFLO state in
 1D  has ranged from applying  Bethe ansatz
\cite{Orso2007,Hu2007,Kakashvili2009},  
density-matrix renormalization group \cite{Feiguin,HM2010,MolinaPRLComment},
quantum Monte Carlo~\cite{Batrouni},   tight binding models
\cite{BakhtiariPRL2008,LuscherPRA2008},  and  self-consistent mean-field
solutions to the gap equation \cite{LiuPRA2007,LiuPRA2008,Sun2011,LuPRL2012,SunBolech}.
Many of these methods only strictly apply to infinite
1D systems, further relying  on the uncontrolled local density approximation (LDA) to account
for the effects of the trapping potential that is omnipresent in ultracold atomic experiments.  

In Ref.~\cite{SheehyPRA2015}
it was shown that a simple variational BCS-type wave function that involved pairing of harmonic-oscillator
states could account for pairing correlations in a balanced trapped 1D gas without the necessity of 
invoking the LDA.  Here, we propose a similar wave function for the imbalanced case that incorporates
FFLO pairing correlations in a natural way.  The wave function is:
%
%
\begin{equation}
\label{FFLO wave function}
|\Psi\rangle =\prod_{m=0}^{q-1}\hat{c}^{\dagger}_{m\uparrow}
\prod_{n=q}^{\infty}\left(u_{n}+v_{n}\hat{c}^{\dagger}_{n\uparrow}\hat{c}^{\dagger}_{n-q\downarrow}\right)|{\rm vac}\rangle,
\end{equation} where the quantum numbers $(m, n)$ label the discrete
energy levels of the trapped gas and the arrow ($\uparrow,\downarrow$)
represents the atomic hyperfine state.   In the following, the trapping
potential will be taken to be harmonic.  

This wave function consists of two product factors 
acting on the vacuum, with the rightmost factor  corresponding, physically, to the presence of
 imbalanced pairing correlations (characterized by the variational 
parameters $u_n$ and $v_n$)  between a spin-$\uparrow$ fermion at level $n$ 
 and a spin-$\downarrow$ fermion
at level $n-q$, as illustrated in Fig.~\ref{fig1}. The left most factor corresponds to 
a central filled core of unpaired spin-$\uparrow$ fermions.  This variational wave function  has a fixed
total population imbalance (or magnetization) equal to $ q = N_{\uparrow}-N_{\downarrow}$.  In addition,
as we shall show below, the real-space local pairing amplitude associated with $|\Psi\rangle$ exhibits
the oscillatory real-space pairing correlations expected for an FFLO state.  For these reasons, we propose that this is a natural
variational wave function to study the FFLO state of  a 1D trapped Fermi system.  


This paper is organized as follows:  In Sec.~\ref{sec:vare},
we derive the  variational energy equation
using Eq.~\eqref{FFLO wave function} in the harmonic oscillator basis
of the trapping potential.  Note that we work only in the zero-temperature limit, studying
ground-state properties.  
 We further discuss the necessity  of
allowing the oscillator length scales associated with the
single-particle basis to differ from that of the harmonic potential.
The energy is then directly  minimized numerically, without resorting
to explicitly  finding the Euler-Lagrange equations.  The results of
this minimization are  presented and analyzed in Sec.~\ref{Results}, in which
the  density profiles and pairing amplitude are shown, as well as
density-density correlations.   
In Sec.~\ref{Conclusion} we make brief concluding remarks.  

Before proceeding to our calculations, we conclude this section by
describing our main results.  We find that an  FFLO state of imbalanced Fermi gases, 
described by Eq.~(\ref{FFLO wave function}), is stable for sufficiently small magnetization
($q$), with a spatially modulated local pairing amplitude that is analogous to the
LO phase of an infinite imbalanced gas, possessing nodes at which $|\Delta(x)|$ vanishes.  
However, we find that the oscillatory pairing amplitude does
not leave any appreciable signature in the local density or local magnetization (density
difference, $m(x)=n_{\uparrow}(x)-n_{\downarrow}(x)$), such as a local
increase in $m(x)$ near the nodes of $|\Delta(x)|$.  
Our results for these densities 
agree qualitatively with experiment and theory based on combining the Bethe ansatz with
the LDA~\cite{LiaoNature2010}, showing an imbalanced central region 
and a paired (balanced) outer region at the edges of the cloud.  With increasing $q$,
the FFLO phase becomes unstable (as in three-dimensional imbalanced Fermi gases) to
an unpaired Fermi gas phase.  



\section{Variational  Energy}
\label{sec:vare}
\label{Variational  Energy} 
We study a one-dimensional imbalanced Fermi gas, a system that has been achieved experimentally using an optical lattice potential to
create an array of weakly-coupled tubes~\cite{LiaoNature2010,Revelle2016}.  In the limit of large optical lattice depth, it is approximately 
valid to neglect any intertube coupling and we therefore study a single tube with Hamiltonian  
\begin{align}
\label{Hamiltonian}
H=&\sum_{\sigma}\int_{-\infty}^\infty {\rm d}x\, \hat{\Psi}^{\dagger}_{\sigma}(x)\left[-\frac{\hbar^{2}}{2m}\frac{d^2}{dx^2}+V(x)\right]\hat{\Psi}^{}_{\sigma}(x)\nonumber\\&
+\lambda \int_{-\infty}^\infty
 {\rm d}x\, \hat{\Psi}^{\dagger}_{\uparrow}(x)\hat{\Psi}^{\dagger}_{\downarrow}(x)\hat{\Psi}^{}_{\downarrow}(x)\hat{\Psi}^{}_{\uparrow}(x),
\end{align}
where $V(x)=\frac{1}{2}m\omega^{2}x^{2}$ is the harmonic trapping potential characterized by 
the trap frequency $\omega$, and $\lambda=-2\hbar^{2}/(m a_{1{\rm D}})$ with $a_{1{\rm D}}$ being the one-dimensional scattering length \cite{OlshaniiPRL1998}.   The field operators can be expanded in terms of mode operators $\hat{c}_{n\sigma}$ as
\begin{equation}
\hat{\Psi}_{\sigma}(x)=\sum_{n=0}^{\infty}\psi_{n\sigma}(x)\hat{c}_{n\sigma}
\end{equation}
and similarly for $\hat{\Psi}_{\sigma}^\dagger(x)$.
The single-particle states are taken to be harmonic oscillator wave functions:
\begin{equation}
\label{single-particle states}
\psi^{}_{n\sigma}(x)=\frac{1}{\sqrt{2^{n}n!a_{\sigma}\sqrt{\pi}}} e^{-x^{2}/(2a^{2}_{\sigma})}H_{n}(x/a_{\sigma}),
\end{equation}
where $H_{n}(x)$ are the Hermite polynomials and  $a_{\sigma}$  are effective  oscillator length scales.

  Crucially, we include $a_{\sigma}$ in the set of variational parameters.  Thus, in the presence of interactions,
they are in general  different from the natural oscillator length $a=\sqrt{\hbar/(m\omega)}$,
which is determined by the trapping potential.   In Ref.~\cite{SheehyPRA2015} the 
balanced case was studied using a similar variational wave function.  It was found that 
including this additional variational parameter was necessary to allow the cloud size to decrease
with increasing strength of attraction (which is what is expected on physical grounds and is 
 observed experimentally).  

Similarly, in the  imbalanced case, we  allow
the oscillator lengths to be variational parameters to obtain
 realistic density profiles.  As discussed below, we find, in the imbalanced regime, that
the optimal (minimal energy) values of the spin-$\uparrow$ and spin-$\downarrow$ oscillator lengths 
generally satisfy (for  attractive interactions) $a_{\sigma}<a$ and $a_{\uparrow}< a_{\downarrow}$. 
This imples that interaction effects cause the two fermion species to each feel an effective
trapping potential with a frequency that is  larger than the actual trap frequency,
with the majority species trap frequency slightly larger than that of the minority species.  



\begin{figure*}%
\centering
\begin{minipage}{0.25\textwidth}%
\includegraphics[width=\textwidth]{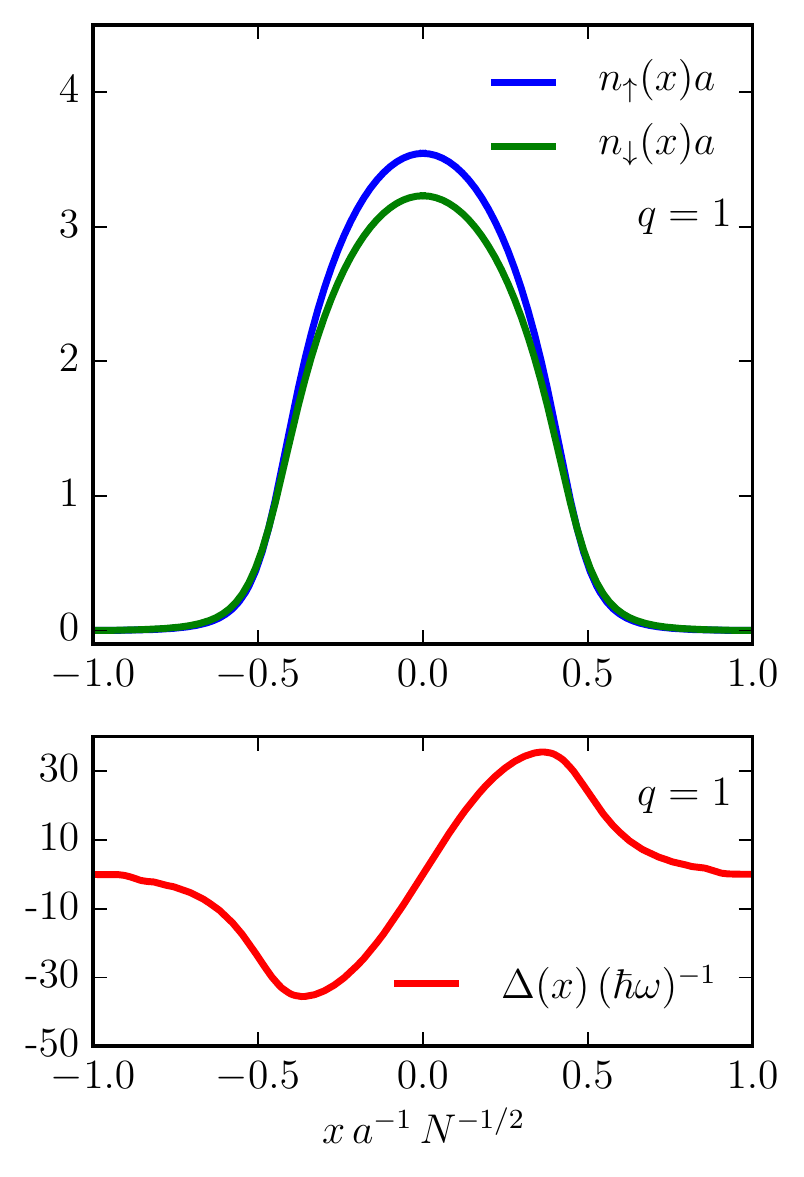}
\end{minipage}%
\hspace{-0.2cm}
\begin{minipage}{0.25\textwidth}%
\includegraphics[width=\textwidth]{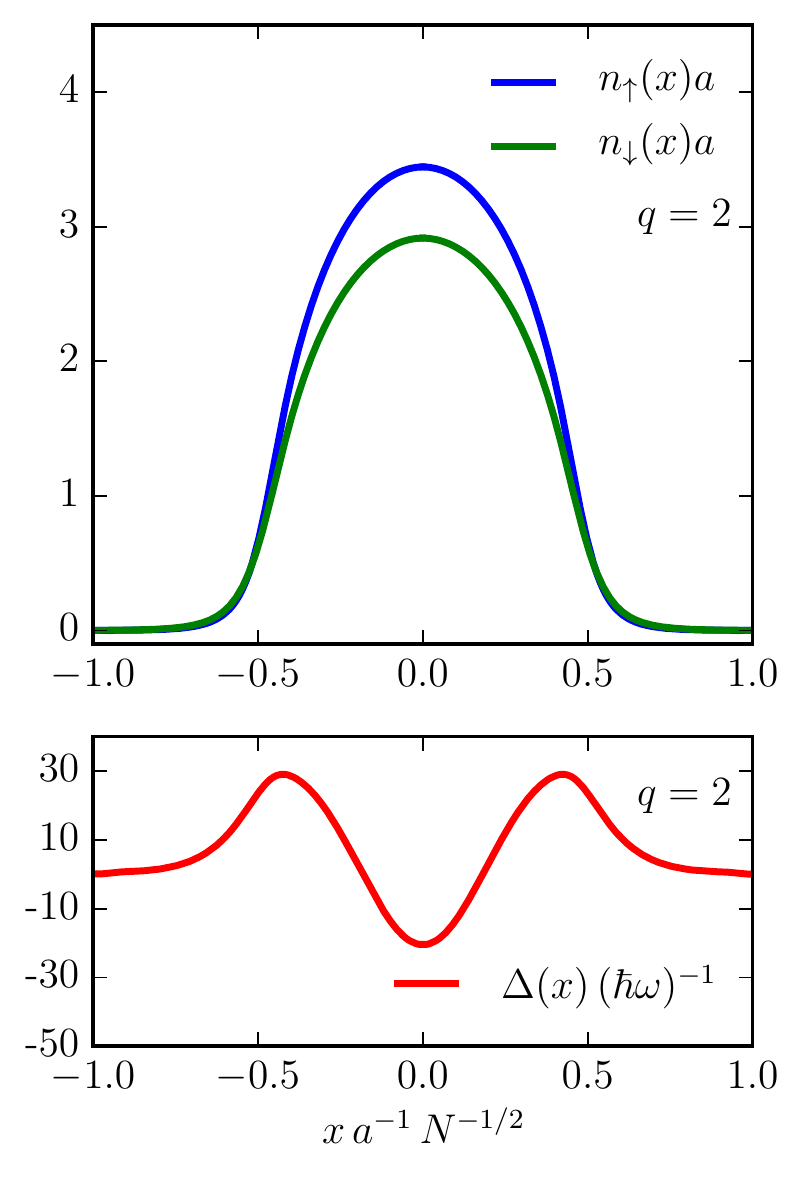}
\end{minipage}%
\hspace{-0.2cm}
\begin{minipage}{0.25\textwidth}%
\includegraphics[width=\textwidth]{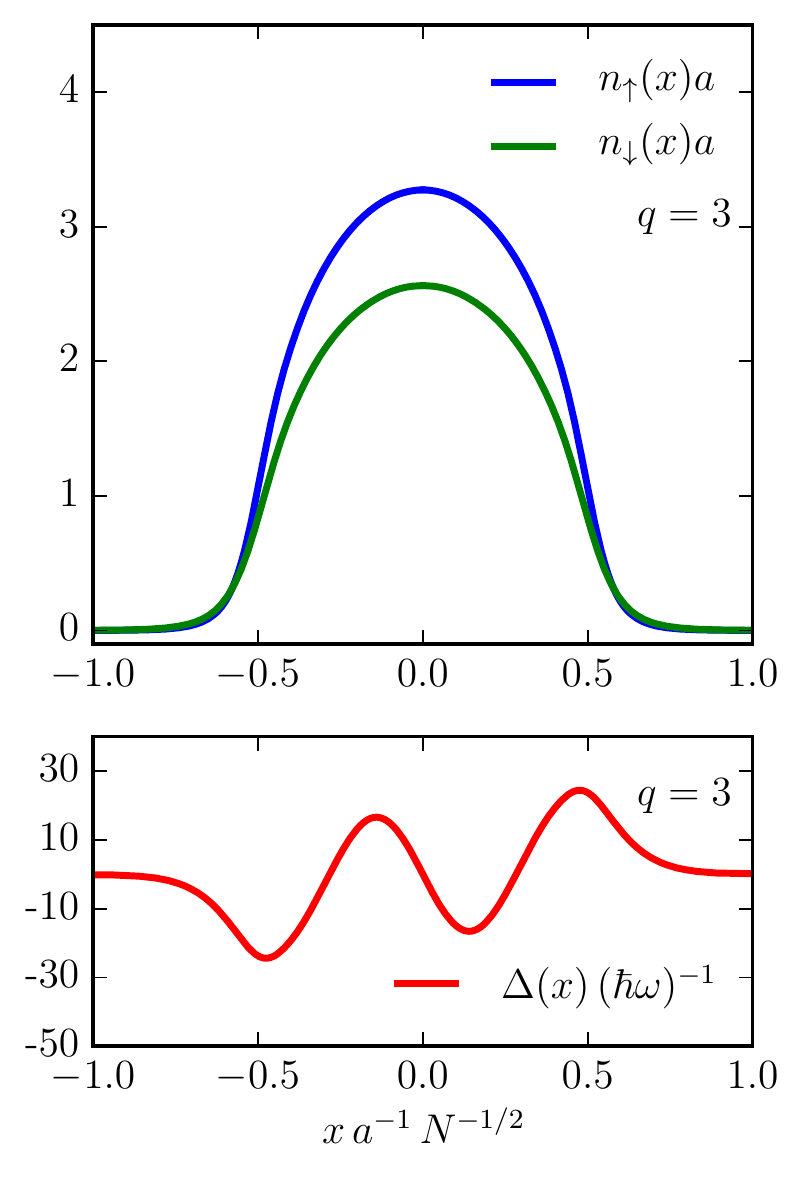}
\end{minipage}%
\hspace{-0.2cm}
\begin{minipage}{0.25\textwidth}%
\includegraphics[width=\textwidth]{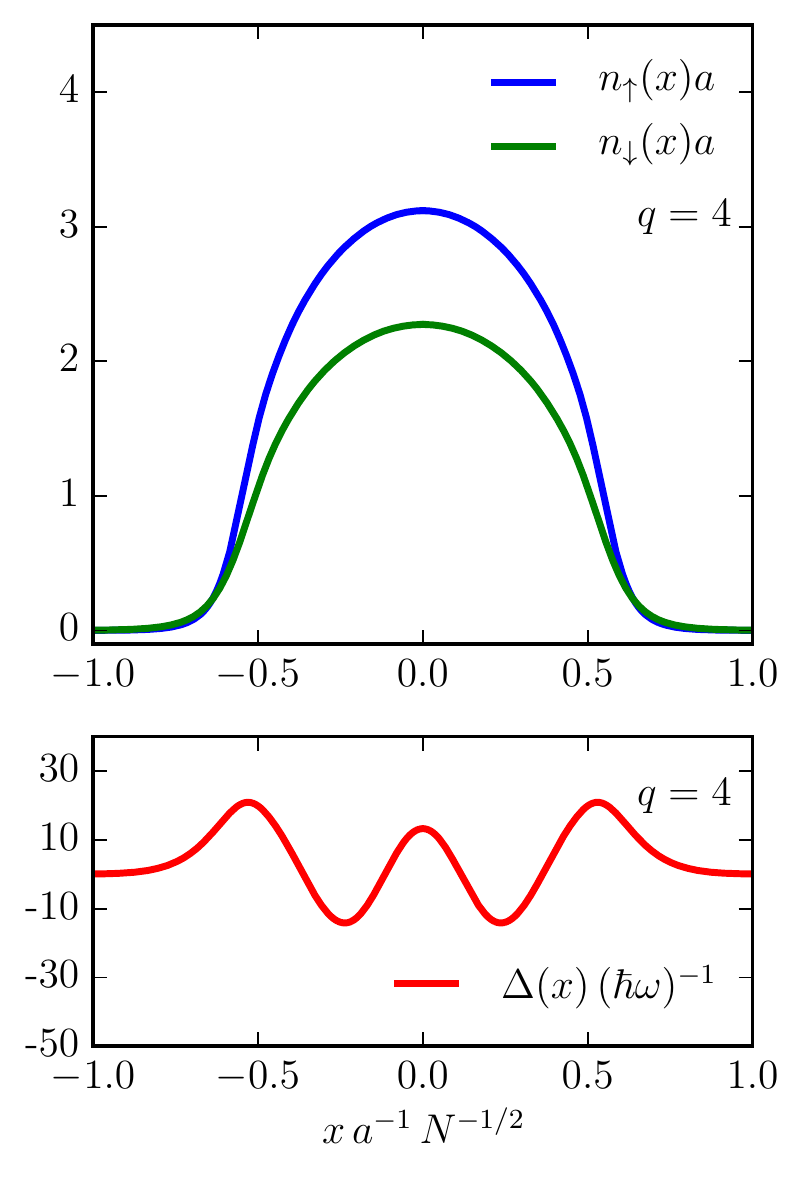}
\end{minipage}%
\caption{(Color Online)  The top row shows the spin-resolved densities, given by Eqs.~\eqref{densities}, of a trapped strongly interacting imbalanced
fermion gas for various  magnetizations
$q = M=N_{\uparrow}-N_{\downarrow}$.  Note that $n_\uparrow(x)\geq n_\downarrow(x)$ always.
 Here,  $a=\sqrt{\hbar/(m\omega)}$ is 
the harmonic oscillator length scale of the
trapping potential.  The
total particle number $N=N_{\uparrow}+N_{\downarrow}\approx 25$.  As
the polarization $P=q/N$ increases from about $P=0.02$ for $q=1$ to
$P=0.08$ for $q=4$ the size of the central locally magnetized region
of the could increases.  The bottom row shows the local pairing
amplitude Eq.~\eqref{local pairing} for the respective values of $q$.
For each additional unpaired spin a node appears in $\Delta(x)$. This
oscillatory structure is a signature of the FFLO state.}
 \label{fig2}%
\end{figure*}

With $a_{\sigma}\neq a$ the single-particle states,
Eq.~\eqref{single-particle states}, are no longer eigenstates of the
kinetic energy term of the Hamiltonian, Eq.~\eqref{Hamiltonian}. 
 However, it is still straightforward to use
the properties of the harmonic oscillator wavefunctions to express the Hamiltonian in terms of the 
mode operators $\hat{c}_{n\sigma}$.  In performing this step, it is 
  convenient to express the trapping potential as
$V(x)=\frac{1}{2}m\omega^{2}_{\sigma}x^{2}+\frac{1}{2}m(\omega^{2}-\omega^{2}_{\sigma})x^{2}$
with $\omega_\sigma = \hbar/(m a_\sigma^2)$ the effective trap associated with the variational oscillator lengths. In terms of mode
operators the Hamiltonian can then be expressed as
\begin{align}
\label{Hamiltonian 2}
H=&\sum_{n,\sigma}\epsilon_{n\sigma}\hat{c}^{\dagger}_{n\sigma}\hat{c}^{}_{n\sigma}+\sum_{\{n_{i}\}}\lambda^{\uparrow,\downarrow}_{n_{1},n_{2},n_{3},n_{4}}\hat{c}^{\dagger}_{n_{1}\uparrow}\hat{c}^{\dagger}_{n_{2}\downarrow}\hat{c}^{}_{n_{3}\downarrow}\hat{c}^{}_{n_{4}\uparrow},
\end{align}
where $\epsilon_{n\sigma}=\frac{\hbar \omega}{2} (n+1/2)(\ell^{2}_{\sigma}+\ell^{-2}_{\sigma})$ with $\ell_{\sigma} = a_{\sigma}/a$, and
\begin{equation}
\label{coupling matrix elements}
\lambda^{\uparrow,\downarrow}_{n_{1},n_{2},n_{3},n_{4}}=\lambda \int_{-\infty}^\infty {\rm d}x\, \psi^{*}_{n_{1}\uparrow}(x)\psi^{*}_{n_{2}\downarrow}(x)\psi^{}_{n_{3}\downarrow}(x)\psi^{}_{n_{4}\uparrow}(x).
\end{equation}
In Eq.~(\ref{Hamiltonian 2}), the summations over the integers $n$ and $n_i$ are from $0$ to $\infty$.  
A convenient expression for Eq.~\eqref{coupling matrix elements}, expressed in terms of a sum over integers,
 is given in the Appendix.  Note, in Eq.~(\ref{Hamiltonian 2})  we have dropped  off-diagonal terms in the kinetic energy, 
as our variational  wave function, Eq.~\eqref{FFLO wave function}, does not connect states having different principal quantum numbers, i.e., $\langle \Psi|\hat{c}^{\dagger}_{n\sigma}\hat{c}_{n'\sigma}|\Psi\rangle=0$ for $n\neq n'$.


Our variational wave function, Eq.~(\ref{FFLO wave function}), has a definite value of the magnetization 
$q = M = N_\uparrow - N_\downarrow$ but an uncertain value of the 
total particle number (as does the usual BCS wave function).  Therefore, we proceed by minimizing the grand-canonical energy 
\begin{align}
{\cal E}=\langle H\rangle-\mu
\langle\hat{N}\rangle,  
\end{align}
with the chemical potential $\mu$, a Lagrange multiplier, that fixes the average total particle number.   
%
Upon evaluating the expectation value and dropping irrelevant constant terms we find for the grand-canonical energy: 
\begin{align}
\label{FFLO expectation value} &{\cal
E}(\{v_{n},u_{n}\},\ell_{\sigma})= \frac{\hbar
\omega}{4}(\ell^{2}_{\uparrow}+\ell^{-2}_{\uparrow})q^{2}\\
&+\sum_{n=q}^{\infty}\big[\epsilon_{n\uparrow}+\epsilon_{n-q\downarrow}-2\mu+\bar{\bar{\lambda}}^{\uparrow,\downarrow}_{n}(q)\big]|v^{}_{n}|^{2}\nonumber\\
&+\sum_{n=q}^{\infty}\sum_{n'=0}^{\infty}\lambda^{\uparrow,\downarrow}_{n,n'}|v_{n}|^{2}|v_{n'+q}|^{2}+\sum_{n,n'=q}^{\infty}\bar{\lambda}^{\uparrow,\downarrow}_{n,n'}(q)u^{*}_{n}v^{}_{n}u^{}_{n'}v^{*}_{n'},
\nonumber
\end{align}  
where for notational convenience we have defined
\begin{subequations}
\label{Eq:specialcases}
\begin{eqnarray}
\lambda^{\uparrow,\downarrow}_{n,n'}&\equiv &
\lambda^{\uparrow,\downarrow}_{n,n',n',n},
\\
\bar{\lambda}^{\uparrow,\downarrow}_{n,n'}(q)&\equiv &
\lambda^{\uparrow,\downarrow}_{n,n-q,n'-q,n'},
\\
\bar{\bar{\lambda}}^{\uparrow,\downarrow}_{n}(q)&\equiv &\sum_{m=0}^{q-1}\sum_{m'=0}^{q-1}\lambda^{\uparrow,\downarrow}_{m,n-q,n-q,m'}.
\end{eqnarray}
\end{subequations}
We see that the coupling function appears in three places in the grand-canonical energy function.
The term containing $\bar{\bar{\lambda}}^{\uparrow,\downarrow}_{n}$
 in the second line of Eq.~(\ref{FFLO expectation value}) describes a Hartree interaction between
the paired and unpaired atoms.  In the third line of  Eq.~(\ref{FFLO expectation value}), the term 
containing  $\lambda^{\uparrow,\downarrow}_{n,n'}$ describes the pairing interaction, and the term
containing $\bar{\lambda}^{\uparrow,\downarrow}_{n,n'}(q)$ describes the Hartree interaction among
the paired atoms.

 Instead of further proceeding  to obtain the Euler-Lagrange
equations  from Eq.~\eqref{FFLO expectation value} we will  seek to
numerically minimize the ground state energy directly with respect to
the parameters $v_{n}$, $u_{n}$, and $a_{\sigma}$ (or
$\ell_{\sigma}$), with the chemical potential adjusted to yield the
correct particle number via 
\begin{equation}
N = \langle \hat{N}\rangle = - \frac{\partial {\cal
E}}{\partial \mu}.
\end{equation}
 Once the optimum values of the parameters are known, all other ground
state properties can be obtained, e.g., densities, pairing amplitude,
etc.  To this end,  with the constraint $|v_{n}|^{2}+|u_{n}|^{2}=1$,
it will be useful to parameterize $v_{n}=\cos(\theta_{n})$ and
$u_{n}=\sin(\theta_{n})$. The ground state energy can then be written
as
\begin{align}
\label{FFLO expectation value 2} &{\cal
E}(\{\theta_{n}\},\ell_{\sigma})=\frac{\hbar
\omega}{4}(\ell^{2}_{\uparrow}+\ell^{-2}_{\uparrow})q^{2}\nonumber\\&+\sum_{n=q}^{\infty}\big[\epsilon_{n\uparrow}
+\epsilon_{n-q\downarrow}-2\mu+\bar{\bar{\lambda}}^{\uparrow,\downarrow}_{n}(q)\big]\cos^{2}(\theta_{n})\nonumber
\\&+\sum_{n=q}^{\infty}\sum_{n'=0}^{\infty}\lambda^{\uparrow,\downarrow}_{n,n'}\cos^{2}(\theta_{n})\cos^{2}(\theta_{n'+q})\nonumber
\\&+\sum_{n,n'=q}^{\infty}\bar{\lambda}^{\uparrow,\downarrow}_{n,n'}(q)\sin(\theta_{n})\cos(\theta_{n})\sin(\theta_{n'})\cos(\theta_{n'}). 
\end{align}
To numerically minimize Eq.~\eqref{FFLO expectation value 2} with respect to the set of angles $\theta_{n}$ and normalized  oscillator lengths $\ell_{\sigma}$ an upper cutoff to the number of  oscillators levels included has to be set. For each coupling strength and  particle number we choose a large enough cutoff  such that the final results are insensitive to this value.  In the next section, we describe our results.  
\section{Results}

\begin{figure}
\includegraphics[scale=0.9]{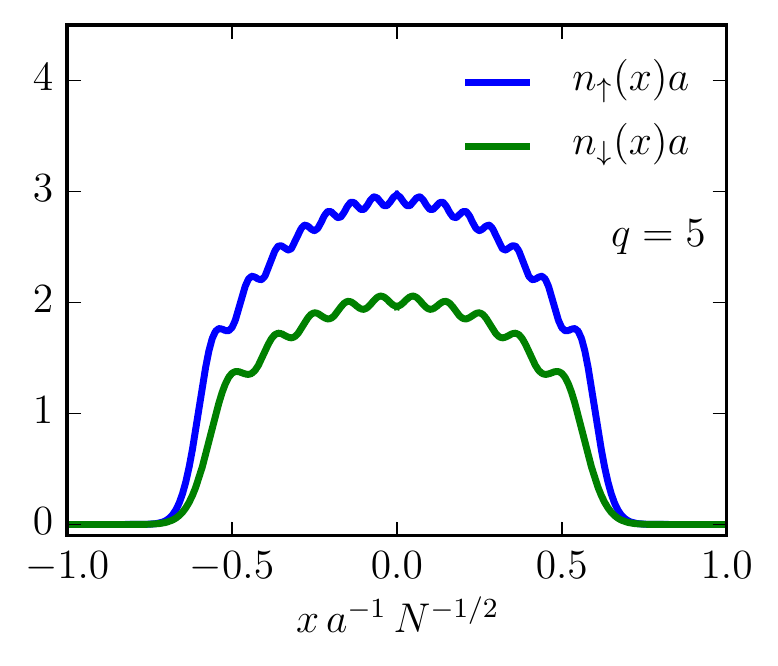}%
\caption{(Color Online) Density profiles for spin-$\uparrow$ and spin-$\downarrow$ for $q=5$ and $N=25$, in the regime of $P>P_c$ where
the unpaired state has lower energy than the paired state, so that $\Delta(x) = 0$.  Note that $n_\uparrow(x)>n_\downarrow(x)$ for all $x$. }\label{fig6}
\end{figure}

\begin{figure*}%
\centering
\begin{minipage}{0.25\textwidth}%
\includegraphics[width=\textwidth]{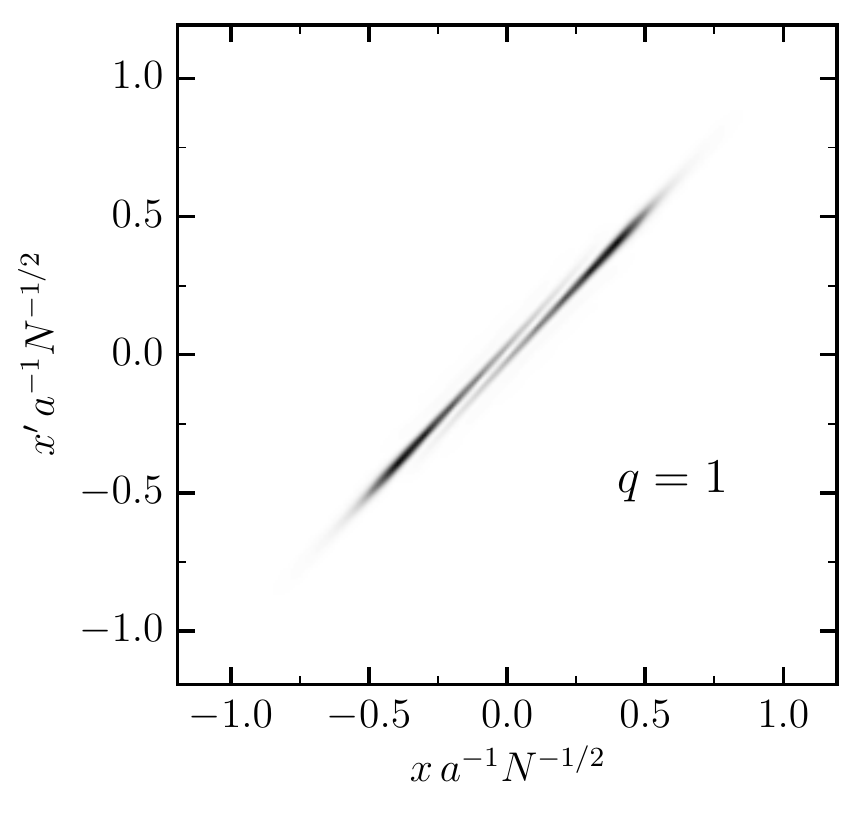}
\end{minipage}%
\hspace{-0.2cm}
\begin{minipage}{0.25\textwidth}%
\includegraphics[width=\textwidth]{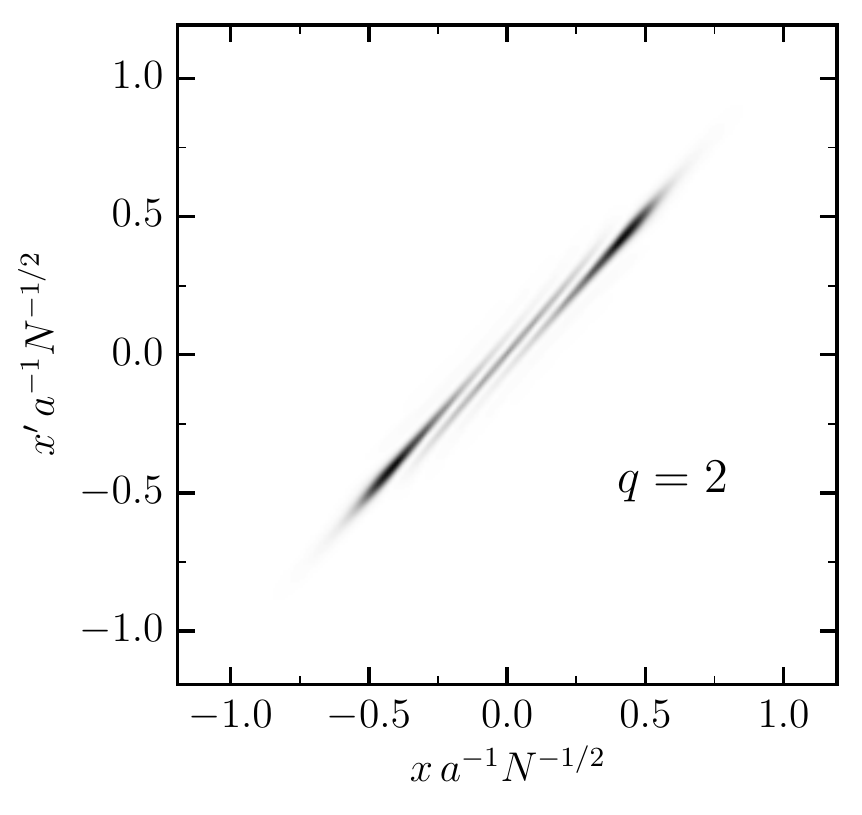}
\end{minipage}%
\hspace{-0.2cm}
\begin{minipage}{0.25\textwidth}%
\includegraphics[width=\textwidth]{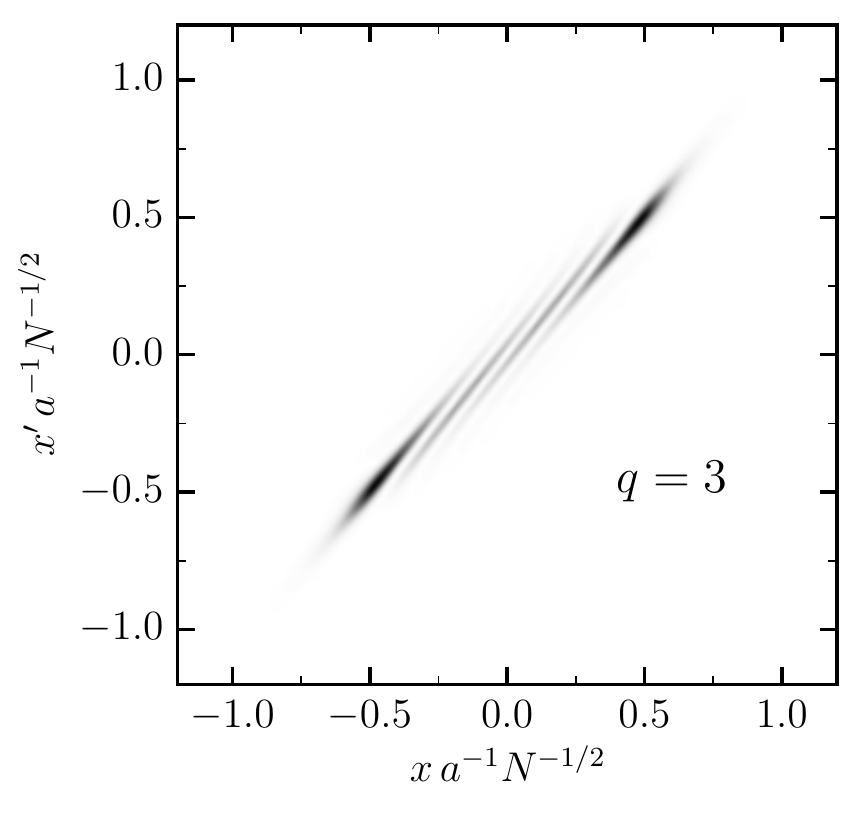}
\end{minipage}%
\hspace{-0.2cm}
\begin{minipage}{0.25\textwidth}%
\includegraphics[width=\textwidth]{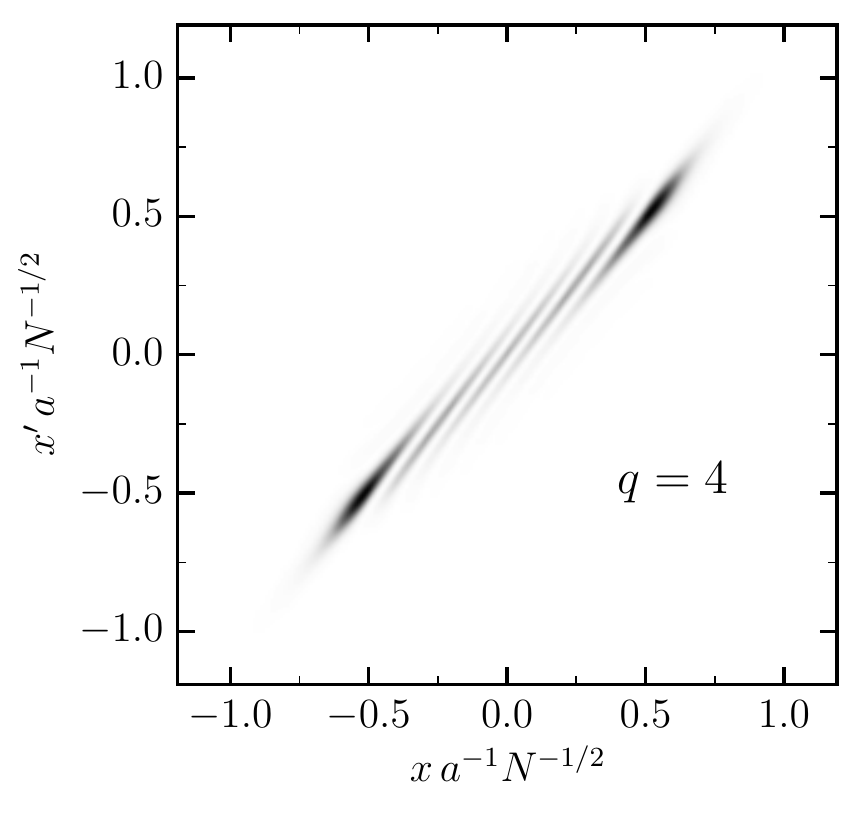}
\end{minipage}%
\caption{Real space (in situ) density-density correlations Eq.~\eqref{in situ
density-density} for various values of the  imbalance $q$. The two
large ``peaks'' in each correspond to the regions shown in
Fig.~\ref{fig2} where the spin-$\uparrow$ and spin-$\downarrow$
densities are approximately equal. The oscillations between these
regions are related to FFLO correlations. Cross-sections along the
diagonal of each of these plots are shown in Fig.~\ref{fig4}}
 \label{fig3}%
\end{figure*}

\label{Results}
The spin-resolved local 
densities $n_{\sigma}(x)=\langle \hat{\Psi}^{\dagger}_{\sigma}(x)\hat{\Psi}^{}_{\sigma}(x)\rangle $ are typical observables in ultracold atomic systems,
where the expectation value is   
with respect to Eq.~\eqref{FFLO wave function}.  The local densities are:
\begin{subequations}\label{densities}
\begin{align}
\label{densitiesa}
n_{\uparrow}(x)& =\sum_{n=0}^{q-1}|\psi_{n\uparrow}(x)|^{2}+\sum_{n=q}^{\infty}|\psi_{n\uparrow}(x)|^{2}|v_{n}|^{2},
\\n_{\downarrow}(x)&=\sum_{n=q}^{\infty}|\psi_{n-q\downarrow}(x)|^{2}|v_{n}|^{2},\label{densitiesb}
\end{align}
\end{subequations}
with the first (second) term of Eq.~(\ref{densitiesa}) being the density of unpaired (paired) majority species (i.e., spin-$\uparrow$) 
fermions.  Within our variational 
ansatz, all spins-$\downarrow$ are pairied, as seen in Eq.~(\ref{densitiesb}).  Note that $n_\uparrow(x)$ and $n_\downarrow(x)$ depend
on the variational parameters $v_n$ but also on $a_\uparrow$ and $a_\downarrow$ via the oscillator wavefunctions.  
Similarly, the local
 pairing amplitude $\Delta(x)=-\lambda\langle \hat{\Psi}^{}_{\downarrow}(x)\hat{\Psi}^{}_{\uparrow}(x)\rangle $ is
\begin{equation}
\label{local pairing}
\Delta(x)=-\lambda\sum_{n=q}^{\infty}\psi_{n-q\downarrow}(x)\psi_{n\uparrow}(x)u_{n}v_{n}.
\end{equation}
Although it is not directly measurable, the magnitude of $\Delta(x)$ measures the strength of the local pairing (and the local single-particle gap).

The top row of Fig.~\ref{fig2} shows the obtained local density
profiles for various values of the total magnetization $q$, while
the bottom row shows the local pairing amplitude that, as expected for an FFLO state,
is oscillatory in real space.  The other system parameters are
 $\lambda =-20\, \hbar\omega  a$ and $N\approx 25$, so that the dimensionless
interaction parameter $N a^{2}_{\rm 1D}/a^{2}\approx 0.25$, which
corresponds to the strongly interacting regime \cite{Orso2007} and the
approximate value in recent experiments \cite{LiaoNature2010}.  As can
be seen, the central region of the cloud is magnetized, while at the
edges of the cloud the two densities remain approximately equal. The
size of this imbalanced region increases with increasing polarization
$P=q/N$.  

The same behavior for the density profiles, i.e., an imbalanced central 
region and balanced outer region, was seen in the experimental
results of Ref.\cite{LiaoNature2010} and found to be consistent
with a theoretical analysis based on combining Bethe ansatz with the 
LDA.  Within such a theoretical description, the observed 
density profiles are interpreted in terms of the cloud being in a locally  imbalanced phase in the center and in a locally
paired phase  at the edges.

\begin{table}[ht]
\centering
\begin{tabular}{c c c c}
\hline
$q$ & $a_{\uparrow}/a$ & $a_{\downarrow}/a$ \\  
\hline
1 & 0.45 & 0.47  \\
2 & 0.47 & 0.51  \\
3 & 0.51 & 0.57  \\
4 & 0.55 & 0.64  \\
\hline
\end{tabular}
\caption{Spin-dependent oscillator lengths  normalized to the oscillator length set by the trapping potential $a$ for various values of the total magnetization $q$.}
\label{Table 1}
\end{table} 

Since our calculations do not make use of the LDA, we instead interpret
our density profiles as arising from the system being in the FFLO state
described by Eq.~(\ref{FFLO wave function}).  To understand why the edges are 
locally paired, we note that the result of minimizing Eq.~(\ref{FFLO expectation value 2})  yields the variational parameters $u_n$ and $v_n$ but also the oscillator
length variational parameters, which we display in  Table \ref{Table 1}.  These results show that the effective oscillator lengths 
($a_\uparrow$ and $a_\downarrow$)  are significantly smaller
than the real system oscillator length, representing the expected contraction of the cloud due to the attractive interactions.  In addition, 
$a_\uparrow<a_\downarrow$, so that the spins-$\downarrow$ contract somewhat less than the spins-$\uparrow$ (with a difference that increases 
with increasing imbalance).
By entering a ground state with $a_\uparrow<a_\downarrow$, our system is effectively allowing the minority
spin-cloud to expand relative to the majority spin-cloud, so that the system is locally balanced (which favors pairing) at the edges of the
cloud, as seen in Fig.~\ref{fig2}.

\begin{figure}
\includegraphics[scale=1]{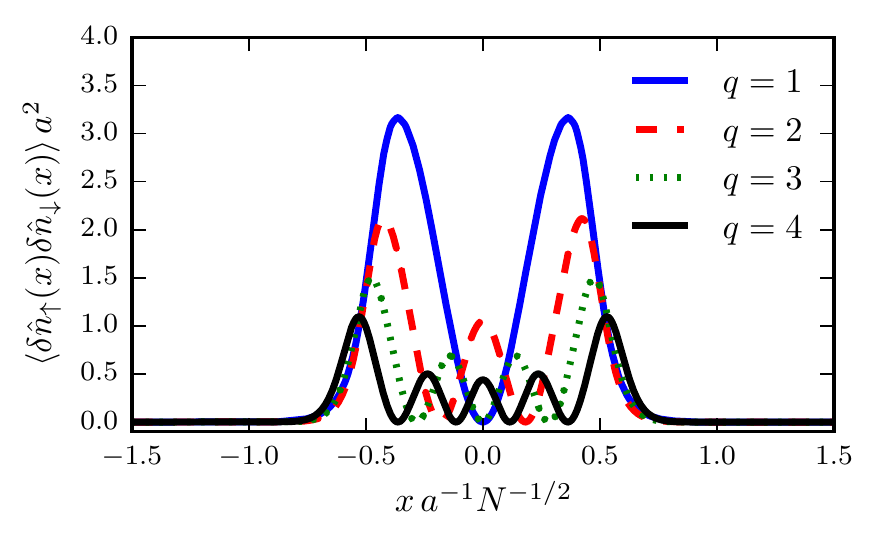}%
\caption{(Color Online) Cross sections of  Fig.~\ref{fig3} along the diagonal $x=x'$
of the in situ density-density correlation function Eq.~\eqref{in situ
density-density}. Note that the $q=1$ case possesses the largest magnitude, and the $q=4$ case the
smallest.  
This observable measures the square of the pairing amplitude
shown in Fig.~\ref{fig2}, directly revealing the  FFLO pairing
oscillations. \label{fig4}}
\end{figure}

Although the pairing amplitude shows oscillatory behavior in real space,
with nodes reminiscent of the LO phase (with the number of nodes equal to $q$),
 no discernable 
signature of the pairing oscillations is reflected in the
spin resolved densities.  We also find that the FFLO state is only stable for sufficiently small $q$, reflecting
a critical polarization $P_{\rm c}$ above which the unpaired Fermi-gas
state becomes energetically favored.  For $\lambda =-20\, \hbar\omega
a$ this occurs  for $q>4$ ($P_{\rm c}=0.16$) with $N=25$ and  $q>2$
($P_{\rm c}=0.04$) with $N=50$. Figure~\ref{fig6} shows the densities of the two spin
states for the case of $q=5$, above the critical polarization. The small wiggles in the
curves are due to the harmonic oscillator wavefunctions comprising the imbalanced
Fermi gas ground state (since, within our theory, this phase is essentially
an imbalanced Fermi gas state).  Henceforth we concentrate on the FFLO phase and leave mapping the phase diagram  for future
work.

\begin{figure*}%
\centering
\begin{minipage}{0.25\textwidth}%
\includegraphics[width=\textwidth]{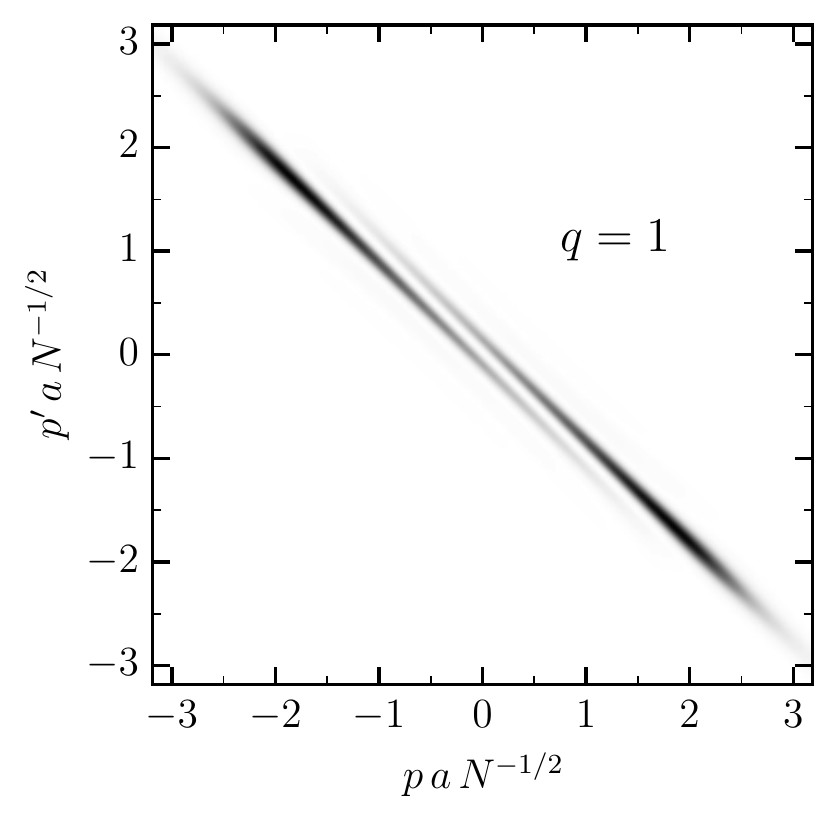}
\end{minipage}%
\hspace{-0.2cm}
\begin{minipage}{0.25\textwidth}%
\includegraphics[width=\textwidth]{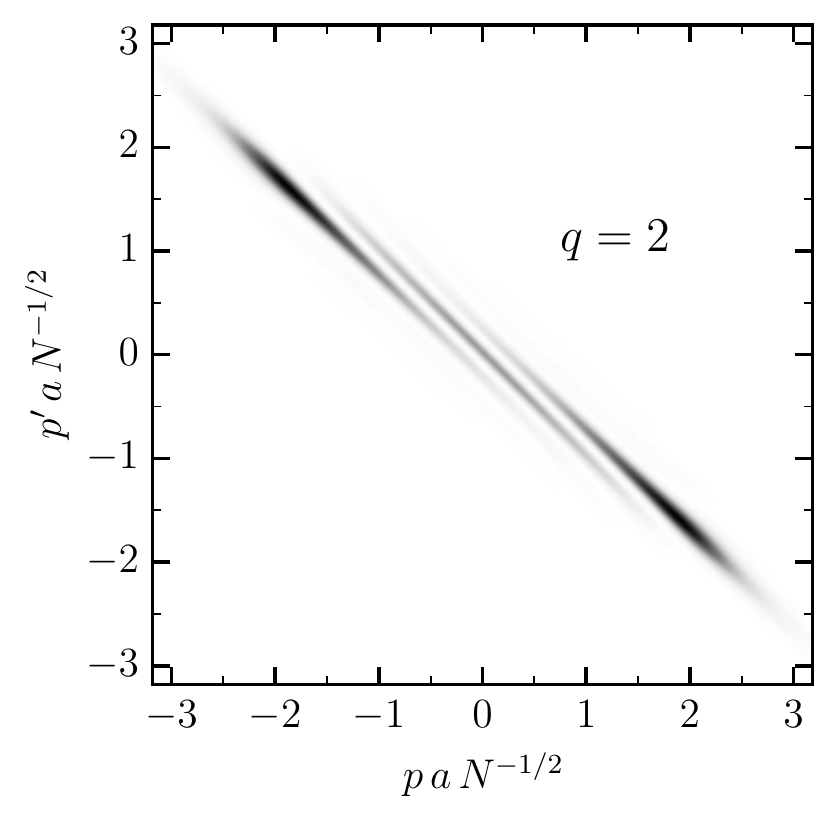}
\end{minipage}%
\hspace{-0.2cm}
\begin{minipage}{0.25\textwidth}%
\includegraphics[width=\textwidth]{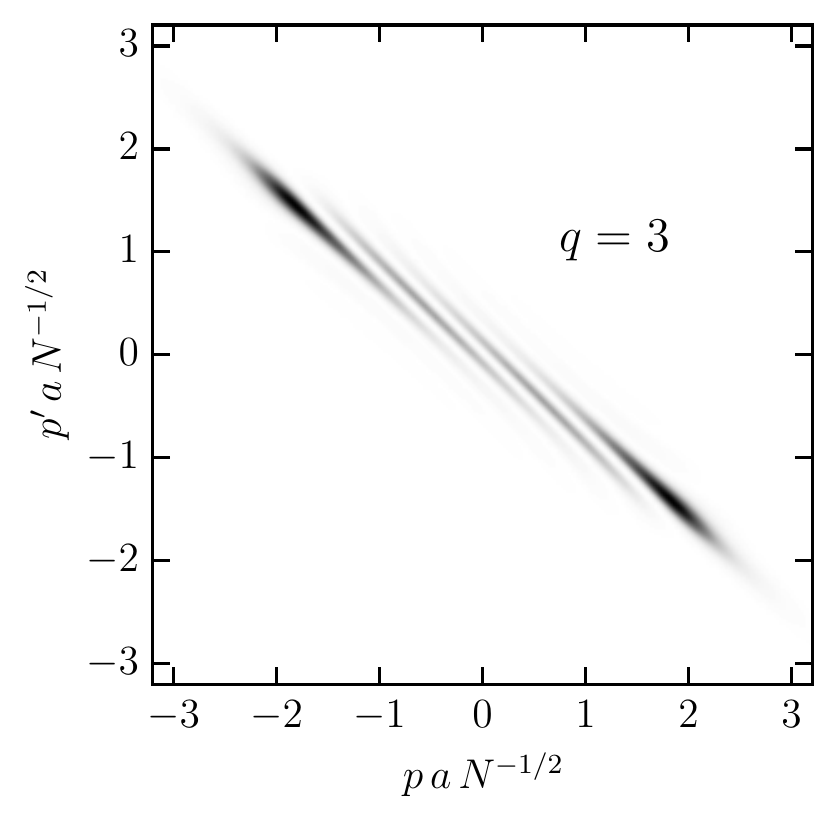}
\end{minipage}%
\hspace{-0.2cm}
\begin{minipage}{0.25\textwidth}%
\includegraphics[width=\textwidth]{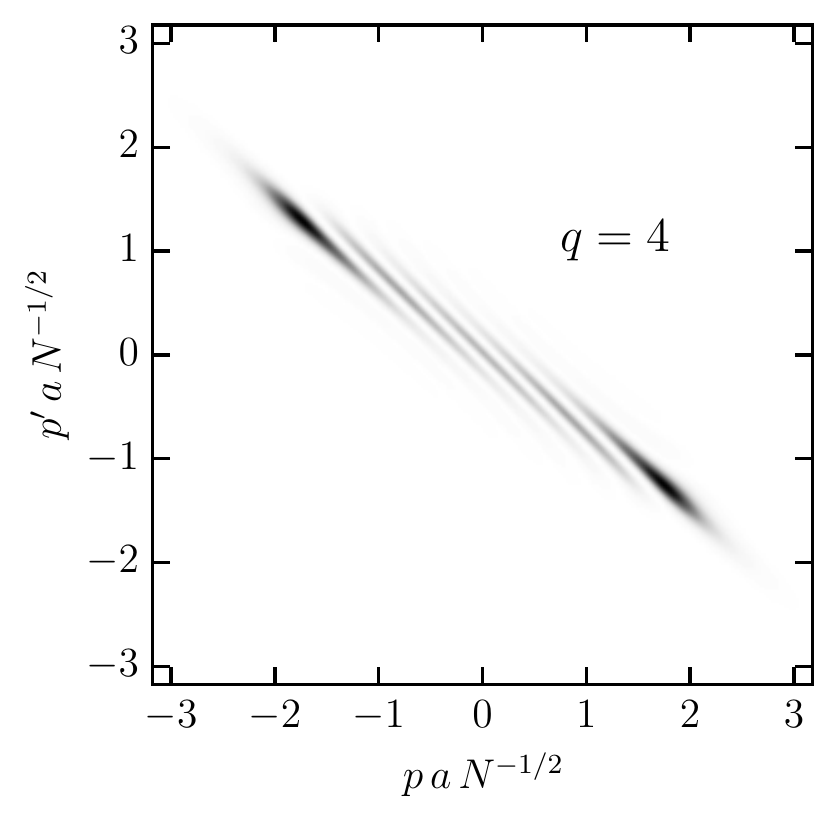}
\end{minipage}%
\caption{Density-density correlations in momentum space (probed by
measuring the real-space density correlations after free expansion~\cite{Altman04,Greiner05}),
Eq.~\eqref{momentum-momentum correlations}, for various imbalances.
The two peaks correspond to Cooper-paired states $(k_{{\rm
F}\uparrow}, -k_{{\rm F}\downarrow})$  and $(-k_{{\rm F}\uparrow},
k_{{\rm F}\downarrow})$ or $(k_{{\rm F}\uparrow}, Q-k_{{\rm
F}\uparrow})$  and $(-k_{{\rm F}\uparrow}, k_{{\rm F}\uparrow}-Q)$,
where $Q$ is the momentum of the FFLO pair.}
 \label{fig5}
\end{figure*}

We now turn to the question of how to directly probe the 1D FFLO state in an experiment via density-density correlations. 
Density-density correlations can be performed after free expansion~\cite{Altman04,Greiner05}
 and in situ \cite{HungNJP}.  We first investigate the latter for our system.
Figure~\ref{fig3} shows the equal-time in situ density correlations defined as
\begin{align}
\label{in situ density-density}
\Pi(x,x')&=\langle \delta \hat{n}_{\uparrow}(x) \delta \hat{n}_{\downarrow}(x) \rangle,\nonumber
\\&=\left|\sum_{n=q}^{\infty}\psi_{n-q\downarrow}(x)\psi_{n\uparrow}(x')u_{n}v_{n}\right|^{2},
\end{align}
where $\delta \hat{n}_{\sigma}(x)=\hat{n}_{\sigma}(x)-\langle \hat{n}_{\sigma}(x) \rangle$, for various imbalances and the same parameters as Fig.~\ref{fig2}.

The two highly  correlated regions  seen in each plot of Fig.~\ref{fig3} correspond to the edges of the cloud where the  spin-$\uparrow$ and spin-$\downarrow$ densities are approximately equal, seen in Fig.~\ref{fig2}. In between these regions the correlations show an oscillatory  behavior.  In fact, along the diagonal $\Pi(x,x)\propto |\Delta(x)|^{2}$.   Thus, the square of the  pairing amplitude can be directly probed, explicitly revealing the FFLO oscillations.  
Figure \ref{fig4} shows  cross-sections of the density-density correlations along $x=x'$ for each plot of Fig.~\ref{fig3}, unambiguously showing the modulated pairing amplitude that is  characteristic  of the FFLO state.

Next we turn to momentum correlations in the trapped FFLO state, which has a similar experimental signature. 
Assuming inter-particle interactions can be neglected during the
expansion \cite{LuPRL2012},  measuring the density after releasing the
trapping potential and allowing the gas to freely expand  probes the
momentum distribution $\langle \hat{n}_{\sigma}(k) \rangle$, where
$\hat{n}_{\sigma}(k) =\hat{c}^{\dagger}_{k\sigma}\hat{c}^{}_{k\sigma}$
and $\hat{c}^{(\dagger)}_{k\sigma}$ annihilates (creates)
single-particle  plane wave  states having momentum $k$.  The noise
correlations of such a measurement are  proportional to the
momentum-space density correlations~\cite{Altman04,Greiner05}. Figure \ref{fig5} shows  the  equal-time momentum-space density correlation function  
\begin{align}
\label{momentum-momentum correlations}
C(k,k')&=\langle \delta\hat{n}_{\uparrow}(k)\delta \hat{n}_{\downarrow}(k') \rangle \nonumber\\&\propto 
\left|\sum_{n=q}^{\infty}\psi_{n-q\downarrow}(k)\psi_{n\uparrow}(k')u_{n}v_{n}\right|^{2},
\end{align}
where $\delta \hat{n}_{\sigma}(k)=\hat{n}_{\sigma}(k)-\langle \hat{n}_{\sigma}(k) \rangle$ and $\psi_{n\sigma}(k)$ is the Fourier transform of the harmonic oscillator states, Eq.~(\ref{single-particle states}),  for various imbalances and the same parameters as Fig.~\ref{fig2}.

As one might expect, the largest correlations are along the
anti-diagonal $k= -k'$.  The two regions of each graph  showing the
highest correlations correspond to momentum states near the Fermi
surfaces (points in one-dimension): $(k_{{\rm F}\uparrow}, -k_{{\rm
F}\downarrow})$  and $(-k_{{\rm F}\uparrow}, k_{{\rm F}\downarrow})$
or in terms of the momentum of the FFLO Cooper-pair $Q$ $(k_{{\rm
F}\uparrow}, Q-k_{{\rm F}\uparrow})$  and $(-k_{{\rm F}\uparrow},
k_{{\rm F}\uparrow}-Q)$. Therefore because of the Fermi surface
mismatch in an imbalanced system the two regions of highest
correlations are off-set from the anti-diagonal by amount
approximately given by the momentum of the Cooper pairs.   Unlike the
in situ density correlations that can give information about the local
pairing amplitude, the momentum correlations  gives information about
the momentum of the Cooper pairs.  Taking together they provide
unambiguous signatures of  two related  predictions of the FFLO
state. 
\section{Conclusion}
\label{Conclusion} Superconductivity, or fermionic superfluidity 
in the presence of a
population imbalance existing  beyond the Chandrasekhar-Clogston
limit has been a topic of interest for many decades. A
long-sought-after candidate, the FFLO state, has been very elusive.
Ultracold atomic gases provide an almost ideal system to produce such
exotic superfluid states because the particle number, interactions,
dimensionally  can all be externally controlled.   In these systems
the inhomogeneity  caused by the trapping potential can have
nontrivial effects. To this end, here we have proposed a ground-state
variational BCS-like wave function for an imbalanced system that
explicitly takes into account the effects of the trapping potential without 
relying on the local density approximation.  

We find that FFLO type pairing
is only stable  for relatively small values of the polarization $P = q/N$, with an imbalanced
unpaired phase stable at larger $P$.  In the small $P$ regime where
the FFLO is stable,  we find density profiles that, while
qualitatively consistent with experiments~\cite{LiaoNature2010}, 
do not show any signature of FFLO pairing correlations in the density.
 However, we find that two-particle
correlations (such as the in situ and free expansion density-density
correlations) show clear experimental signatures of the
local pairing oscillations and finite Cooper pair momentum of an FFLO
superfluid.  

\begin{acknowledgments}
KRP would like acknowledge  support from a Georgia Gwinnett College Seed Fund Grant.   DES acknowledges support from National Science Foundation Grant No. DMR-1151717.
This work of DES was performed in part at the Aspen Center for Physics, which is supported by National Science Foundation grant PHY-1066293.
\end{acknowledgments}
\appendix*
\section{Interaction matrix elements}
In this section we provide simplified expressions for the various interaction matrix elements required for our calculations.  Our 
starting point is the interfermion interaction Eq.~\eqref{coupling matrix elements}:

\begin{equation}
\lambda^{\uparrow,\downarrow}_{n_{1},n_{2},n_{3},n_{4}}=\lambda \int_{-\infty}^\infty {\rm d}x\, \psi^{*}_{n_{1}\uparrow}(x)\psi^{*}_{n_{2}\downarrow}(x)\psi^{}_{n_{3}\downarrow}(x)\psi^{}_{n_{4}\uparrow}(x),
\end{equation}
where 
\begin{equation}
\psi^{}_{n\sigma}(x)=\frac{1}{\sqrt{2^{n}n!a_{\sigma}\sqrt{\pi}}} e^{-z^{2}/(2a^{2}_{\sigma})}H_{n}(x/a_{\sigma}),
\end{equation}
are harmonic oscillator states (or Hermite functions) with spin dependent oscillator lengths $a_{\sigma}$.  Henceforth, we shall drop the integration limits, with the understanding that they are 
always from $-\infty$ to $\infty$.  Plugging in $\psi_{n\sigma}(x)$ leads to: 
\begin{align}
&\lambda^{\uparrow,\downarrow}_{n_{1},n_{2},n_{3},n_{4}}=\frac{\lambda}{\pi a_{\uparrow} a_{\downarrow}}\frac{1}{\sqrt{2^{n_{1}+n_{2}+n_{3}+n_{4}}n_{1}!n_{2}!n_{3}!n_{4}!}}\nonumber\\&\times \int {\rm d}x\, e^{-2x^{2}/\bar{a}^{2}}H_{n_{1}}(x/a_{\uparrow})H_{n_{2}}(x/a_{\downarrow})H_{n_{3}}(x/a_{\downarrow}) \nonumber\\ &\hspace{2.5cm} \times H_{n_{4}}(x/a_{\uparrow}),
\end{align}
where
\begin{equation}
\bar{a}^{2}=\frac{2}{\frac{1}{a^{2}_{\uparrow}}+\frac{1}{a^{2}_{\downarrow}}},
\end{equation}
is the harmonic mean of the square of the oscillator lengths.  Recasting  the integral into dimensionless form by defining $u =  x/\bar{a}$ we have
\begin{align}
&\lambda^{\uparrow,\downarrow}_{n_{1},n_{2},n_{3},n_{4}}=\frac{\lambda\bar{a}}{\pi a_{\uparrow} a_{\downarrow}}\frac{1}{\sqrt{2^{n_{1}+n_{2}+n_{3}+n_{4}}n_{1}!n_{2}!n_{3}!n_{4}!}}\nonumber\\&\times \int {\rm d}u\, e^{-2u^{2}}H_{n_{1}}(u \bar{a}/a_{\uparrow})H_{n_{2}}(u \bar{a}/a_{\downarrow})H_{n_{3}}(u\bar{a}/a_{\downarrow})\nonumber\\ &\hspace{2cm} \times H_{n_{4}}(u\bar{a}/a_{\uparrow}).
\end{align}
Thus we need to evaluate the following integral:
\begin{align}
&I=\int {\rm d}u\, e^{-2u^{2}}H_{n_{1}}(u \bar{a}/a_{\uparrow})H_{n_{2}}(u \bar{a}/a_{\downarrow})H_{n_{3}}(u\bar{a}/a_{\downarrow}) \nonumber\\ &\hspace{2.25cm} \times H_{n_{4}}(u\bar{a}/a_{\uparrow}).
\end{align}
In principle this integral can be express in terms of a Lauricella function of the second kind \cite{Lord1948}, but this is still computationally  inefficient to evaluate.  Thus, we use the Feldheim relation for the product of two Hermite polynomials
\begin{equation}
H_{m}(z)H_{n}(z)=m!n!\sum_{\nu=0}^{{\rm min}(m,n)}\frac{2^{\nu}H_{m+n-2\nu}(z)}{\nu!(m-\nu)!(n-\nu)!},
\end{equation}
to arrive at
\begin{align}
I=&n_{1}!n_{2}!n_{3}!n_{4}!\sum_{\nu_{1}=0}^{{\rm min}(n_{1},n_{4})}\sum_{\nu_{2}=0}^{{\rm min}(n_{2},n_{3})}\frac{2^{\nu_{1}}}{\nu_{1}!(n_{1}-\nu_{1})!(n_{4}-\nu_{1})!}\nonumber\\&\times\frac{2^{\nu_{2}}}{\nu_{2}!(n_{2}-\nu_{2})!(n_{3}-\nu_{2})!}\nonumber\\&\times\int {\rm d}u\, e^{-2u^{2}}H_{n_{1}+n_{4}-2\nu_{1}}(u \bar{a}/a_{\uparrow})H_{n_{2}+n_{3}-2\nu_{2}}(u \bar{a}/a_{\downarrow}).
\end{align}
Next we use the following integral identity~\cite{Lord1948}
\begin{align}
&2^{-1/2}\int {\rm d}u e^{-u^{2}}H_{m}(u\alpha)H_{n}(u\beta)=\nonumber\\&
\begin{cases}
 (-1)^{\frac{1}{2}(m-n)} 2^{m+n-\frac{1}{2}} \alpha^{n}\beta^{m}\Gamma\left(\frac{m+n+1}{2}\right) &\text{if } m+n \text{ is even}  \\
0 &\text{otherwise},
\end{cases}
\end{align}
which holds if $\alpha^{2}+\beta^{2}=1$.  With this, we have 
\begin{widetext} 
\begin{align}
\label{general matrix element result}
\lambda^{\uparrow,\downarrow}_{n_{1},n_{2},n_{3},n_{4}}=&\frac{\lambda\bar{a}}{\pi a^{}_{\uparrow}a^{}_{\downarrow}}\sqrt{\frac{n_{1}!n_{2}!n_{3}!n_{4}!}{2^{n_{1}+n_{2}+n_{3}+n_{4}}}}\sum_{\nu_{1}=0}^{\min(n_{1},n_{4})}\sum_{\nu_{2}=0}^{\min(n_{2},n_{3})}\frac{2^{\nu_{1}}}{\nu_{1}!(n_{1}-\nu_{1})!(n_{4}-\nu_{1})!}\frac{2^{\nu_{2}}}{\nu_{2}!(n_{2}-\nu_{2})!(n_{3}-\nu_{2})!}\nonumber\\ &\times J^{\uparrow,\downarrow}_{n_{1}+n_{4}-2\nu_{1},n_{2}+n_{3}-2\nu_{2}},
\end{align}
where 
\begin{align}
&J^{\uparrow,\downarrow}_{m,n}=
\begin{cases}
\displaystyle
 (-1)^{\frac{1}{2}(m-n)} 2^{m/2+n/2-1/2} \left(\frac{\bar{a}}{a_{\uparrow}}\right)^{n}\left(\frac{\bar{a}}{a_{\downarrow}}\right)^{m}\Gamma\left(\frac{m+n+1}{2}\right) &\text{if } m+n \text{ is even}  \\
0 &\text{otherwise}.
\end{cases}
\end{align}
Thus, we have expressed the integration of a product of four Hermite functions (with distinct oscillator lengths),
 Eq.~(\ref{coupling matrix elements}), as a double sum over integers.

From this general result, further simplification can be achieved for the three special cases that are needed in Eq.~(\ref{Eq:specialcases})
 $\lambda^{\uparrow,\downarrow}_{n,n'}$, $\bar{\lambda}^{\uparrow,\downarrow}_{n,n'}(q)$, and $\bar{\bar{\lambda}}^{\uparrow,\downarrow}_{n}(q)$, 
corresponding to pairing interactions, Hartree interactions among pairs, and Hartree interactions between the pairs and excess spins-$\uparrow$,
respectively. 
\subsection{Pairing interaction: $\lambda^{\uparrow,\downarrow}_{n,n'}\equiv\lambda^{\uparrow,\downarrow}_{n,n',n',n}$}
In this limit Eq.~\eqref{general matrix element result}  simplifies to  
\begin{align}
\lambda^{\uparrow,\downarrow}_{n,n'}=&\frac{\lambda}{\sqrt{2}\pi \bar{a}}\frac{n!}{n'!}(-1)^{n-n'} \left(\frac{\bar{a}^{}}{a_{\uparrow}}\right)^{2n'+1}\left(\frac{\bar{a}^{}}{a_{\downarrow}}\right)^{2n+1}\sum_{\nu_{1}=0}^{n}\frac{(-1)^{-\nu_{1}}\Gamma\left(n+n'-\nu_{1}+\tfrac{1}{2}\right)}{\nu_{1}![(n-\nu_{1})!]^{2}}\left(\frac{a^{}_{\downarrow}}{\bar{a}^{}}\right)^{2\nu_{1}}\nonumber\\&\times {}_{2}F_{1}\left(-n',-n';\tfrac{1}{2}-n-n'+\nu_{1};a^{2}_{\uparrow}/\bar{a}^{2}\right),
\end{align}
where ${}_{2}F_{1}(a,b;c;x)$ is a hypergeometric function.  

\subsection{Hartree interaction among pairs: $\bar{\lambda}^{\uparrow,\downarrow}_{n,n'}(q)\equiv\lambda^{\uparrow,\downarrow}_{n,n-q,n'-q,n'}$}
In this limit Eq.~\eqref{general matrix element result}  simplifies to 
\begin{align}
\bar{\lambda}^{\uparrow,\downarrow}_{n,n'}(q)=&\frac{\lambda(-1)^{q}}{\sqrt{2}\pi\bar{a}}\sqrt{\frac{n!n'!}{(n-q)!(n'-q)!}}\left(\frac{\bar{a}}{a_{\uparrow}}\right)^{n+n'-2q+1}\left(\frac{\bar{a}}{a_{\downarrow}}\right)^{n+n'+1}
\sum_{\nu_{1}=0}^{\min(n,n')}\frac{(-1)^{-\nu_{1}}\Gamma\left(n+n'-q-\nu_{1}+\tfrac{1}{2}\right)}{\nu_{1}!(n-\nu_{1})!(n'-\nu_{1})!}\left(\frac{a^{}_{\downarrow}}{\bar{a}^{}}\right)^{2\nu_{1}}\nonumber\\&\times {}_{2}F_{1}\left(-n+q,-n'+q;\tfrac{1}{2}-n-n'+q+\nu_{1};a^{2}_{\uparrow}/\bar{a}^{2}\right).
\end{align}
\subsection{Hartree interaction between pairs and excess spins-$\uparrow$
: $\bar{\bar{\lambda}}^{\uparrow,\downarrow}_{n}(q)\equiv \sum_{m=0}^{q-1}\sum_{m'=0}^{q-1}\lambda^{\uparrow,\downarrow}_{m,n-q,n-q,m'}$ }
 If  $m+m'$ is even then the summand is
\begin{align}
\lambda^{\uparrow,\downarrow}_{m,n-q,n-q,m'}=&\frac{\lambda (-1)^{\frac{m+m'}{2}+q-n}}{\sqrt{2}\pi \bar{a}}\frac{\sqrt{m!m'!}}{(n-q)!}\left(\frac{\bar{a}}{a_{\uparrow}}\right)^{2(n-q)+1}\left(\frac{\bar{a}}{a_{\downarrow}}\right)^{m+m'+1}\sum_{\nu_{1}=0}^{\min(m,m')}\frac{(-1)^{-\nu_{1}}\Gamma\left(\tfrac{m+m'}{2}+n-q-\nu_{1}+\tfrac{1}{2}\right)}{\nu_{1}!(m-\nu_{1})!(m'-\nu_{1})!}\nonumber\\&\times\left(\frac{a^{}_{\downarrow}}{\bar{a}}\right)^{2\nu_{1}} {}_{2}F_{1}\left(-n+q,-n+q;\tfrac{1}{2}-\tfrac{m+m'}{2}-n+q+\nu_{1};a^{2}_{\uparrow}/\bar{a}^{2}\right)
\end{align}
and  zero otherwise.  For small $q$, the summation required to obtain $\bar{\bar{\lambda}}^{\uparrow,\downarrow}_{n}(q)$ is straightforward to evaluate numerically.
\end{widetext}
The numerical evaluation of these terms is still very challenging, as each of these three cases involve an alternating series whose terms grow exponentially large. Thus the final result amounts to taking the difference of two very large and very close numbers.  High precision libraries had to be used in their numeric evaluation \cite{ARB}.

\end{document}